\documentstyle[12pt,aaspp4,flushrt]{article}

\catcode`\@=11 
\def\@versim#1#2{\vcenter{\offinterlineskip
        \ialign{$\m@th#1\hfil##\hfil$\crcr#2\crcr\sim\crcr } }}
\newcommand{\beq}{\begin{equation}}
\newcommand{\eeq}{\end{equation}}
\def\lsim{\mathrel{\mathpalette\@versim<}}
\def\gsim{\mathrel{\mathpalette\@versim>}}

\def\ch{{\em Chandra} }
\def\mpy{\rm \ M_\odot \ {\rm yr^{-1}}}

\begin{document}
\title{A Dynamical Model for Hot Gas in the Galactic Center}
\author{Eliot Quataert} \affil{ UC Berkeley, Astronomy Department, 501
Campbell Hall, Berkeley, CA 94720; eliot@astron.berkeley.edu}

\medskip

\begin{abstract}

Winds from massive stars supply $\approx 10^{-3} \mpy$ of gas to the
central parsec of the Galactic Center.  Spherically symmetric
hydrodynamic calculations show that $\approx 1 \%$ of this gas, or
$\approx 10^{-5} \mpy$, flows in towards the central massive black
hole Sgr A*; the remaining gas, $\approx 10^{-3} \mpy$, is thermally
driven out of the central star cluster in a wind.  This dynamical
model accounts for the level of diffuse X-ray emission observed in the
Galactic Center by $\ch$ and the extended X-ray source coincident with
Sgr A*; the latter is a direct signature of gas being gravitational
captured by the black hole.

\

\noindent {\it Subject Headings:} Galaxy: center --- accretion,
accretion disks

\end{abstract}

\section{Introduction}

{\ch} observations of the center of the Milky Way reveal diffuse gas
within several parsecs of the central massive black hole, Sgr A*
(Baganoff et al. 2003).  This gas undoubtedly originates from the
interaction of the strong stellar winds produced by the several dozen
massive stars in the central parsec star cluster (e.g., Krabbe et
al. 1991; Najarro et al. 1997).  These stellar winds, and the
associated hot X-ray emitting gas, are believed to be the primary
reservoir of material for accretion onto the central black hole (e.g.,
Melia 1992).

In this paper I present a model for the dynamics of the observed hot
gas on scales of $\sim 0.01-1$ pc in the Galactic Center.  I am
motivated by several considerations.  First, the rate at which Sgr A*
accretes surrounding gas is usually estimated using the Bondi
accretion formula.  The resulting accretion rate, $\approx
10^{-5}-10^{-6} \mpy$ (e.g., Melia 1992; Baganoff et al. 2003), is
much less than the total mass loss rate by stars in the central
parsec, $\approx 10^{-3} \mpy$ (Najarro et al. 1997).  Nearly all of
the mass lost by stars must therefore be driven out of the Galactic
Center in a wind (e.g., Chevalier 1992).  To accurately model the gas
gravitationally captured by Sgr A*, one should also self-consistently
account for the dynamics of the unbound wind.  Moreover, during most
epochs, \ch observes an {extended} X-ray source coincident with Sgr
A*, which has a size $\approx R_B \approx 1''$ (Baganoff et al. 2003),
where $R_{B} \approx GM/c^2_s$ is the Bondi accretion radius for gas
of sound speed $c_s$ around a black hole of mass $M$. In a previous
paper (Quataert 2002; see also Yuan et al. 2002), I argued that this
extended source is due to thermal emission from hot gas at $\sim R_B$,
but I did not present a detailed model for the dynamics of this gas,
nor did I quantitatively model the observed surface brightness
profile.  If the thermal emission interpretation of the extended
source is correct, it would be direct evidence for gas being
gravitationally captured by the central black hole, confirming that
$\sim 10^{-5} \mpy$ is flowing in on scales of $\sim R_B$.  The
significance of this inference motivates a better model for the
dynamics of the hot gas observed by {\em Chandra}.


In the next section I incorporate stellar mass loss as a source term
in the hydrodynamic equations and calculate the dynamics of both
accreting and outflowing gas assuming spherical symmetry.  This
extends previous work on a Galactic Center wind (e.g., Chevalier 1992)
to incorporate the effects of the central black hole.  The spherical
assumption is relatively simplistic since a small number of stars
dominate the mass supply in the central parsec star cluster.  I show,
however, that this model reproduces the \ch observations well.  Melia
and collaborators (Coker \& Melia 1997; Rockefeller et al. 2003) have
presented 3D simulations that address some of the issues considered
here.

\section{The Fate of Stellar Winds}

The interaction of multiple stellar winds leads to shocks which heats
the gas to X-ray emitting temperatures.  The dynamics of the hot gas
can be modeled by incorporating stellar winds as a source of mass and
energy in the equations of hydrodynamics.  Assuming spherical
symmetry, the resulting equations are given by (e.g., Holzer \& Axford
1970)

\beq {\partial \rho \over \partial t} + {1 \over r^2}{\partial \over
\partial r} r^2 \rho v = q(r), \label{mass} \eeq \beq \rho {d v \over
d t} = -{\partial p \over \partial r} - \rho {G M \over r^2} - q(r) v,
\label{mom} \eeq and \beq \rho T {d s \over d t} = q(r)\left[{v^2
\over 2} + {v_w^2 \over 2} - {\gamma \over \gamma - 1} c^2_s\right],
\label{en} \eeq where $\rho$, $v$, $c_s$, and $s$ are the mass
density, radial velocity, isothermal sound speed, and entropy per unit
mass, respectively; $M = 3.6 \times 10^6 M_\odot$ is the mass of the
black hole which dominates the gravity on the scales of interest.  For
the densities and temperatures appropriate to the Galactic Center,
radiative cooling is negligible and so has been dropped in eq. (3).
In eqs. (\ref{mass})-(\ref{en}), $q(r)$ is the stellar mass loss rate
per unit volume and $v_w^2/2$ is the rate of energy injection per unit
mass from stellar winds with velocity $v_w$.  The total rate of mass
injection is given by $\dot M_w = \int 4 \pi r^2 q(r)$.  Incorporating
mass loss as a source term eliminates the need to specify boundary
conditions on the density and temperature of gas at an arbitrary
'fiducial' radius, as is required in Bondi accretion and Parker wind
models.  It should be noted that equations analogous to those above
have been used to study gas flow in several other environments, such
as 'galactic winds' from elliptical galaxies (e.g., Mathews \& Baker
1971) and winds from star clusters without black holes (e.g., Canto et
al. 2000).

The source terms in the above equations are well constrained by IR
spectroscopy of the Galactic Center, which reveals a cluster of
massive stars within $\approx 10$'' of the black hole (e.g., Genzel et
al. 2003).  These include blue supergiants with mass loss rates $\sim
10^{-4} \mpy$ and wind speeds $v_w \approx 600-1000$ km s$^{-1}$
(e.g., Najarro et al. 1997).  The total stellar mass loss is $\approx
10^{-3} \mpy$ and is dominated by IRS 13E which is $\approx 3.5''$
from Sgr A* on the sky.  There is an additional cluster of massive
stars much closer to the black hole, namely those whose orbits have
recently been measured (e.g., Sch\"odel et al. 2002; Ghez et
al. 2003a).  Spectroscopy of one such star, S0-2, suggests that it is a
main sequence O/B star (Ghez et al 2003b; Eisenhauer et al. 2003), in
which case its mass loss rate is probably much smaller than that of
the evolved stars further from the black hole.  In what follows I
neglect mass loss from the closer-in star cluster; further
observations are required to check this assumption.

I solve equations (1)-(3) to determine the fate of hot gas in the
Galactic Center.  I choose model parameters based on the observations
described above; for standard parameters I take $v_w \approx 1000$ km
s$^{-1}$ and $\dot M_w \approx 10^{-3} \mpy$.  The biggest uncertainty
is how to model the spatial distribution of mass loss in this
simplified one-dimensional calculation.  Since the observed mass
losing stars are located several arcsec from the black hole, I set
$q(r) \propto r^{-\eta}$ for $r \ \epsilon \ [2'',10'']$, and $q(r) =
0$ otherwise.  The local mass injection rate is given by $d\dot
M_w/d\ln r \propto r^{-\eta + 3}$ so that $\eta = 0$ corresponds to
mass injection that is concentrated at large radii while $\eta = 3$
corresponds to equal mass injection per decade in radius.  Modest
variations about this choice of $q(r)$ yield similar results to those
described below.

After several sound crossing times, the solution of equations (1)-(3)
settles into a steady state in which gas in the inner region is
captured and flows in towards the black hole, while gas further away
is blown out of the system in a wind.  Figures 1 \& 2 show the steady
state radial velocity, temperature, and density as a function of
distance from the black hole for the above parameters with $\eta = 0,
2, 3$. The total accretion rate through the inner boundary in the
three solutions ranges from $\approx 0.0015-0.03 \ \dot M_w \sim
10^{-5} \mpy$; the majority of the gas, $\approx \dot M_w \approx
10^{-3} \mpy$, is driven out of the central star cluster.  As Fig. 1
shows, the division between inflowing and outflowing gas occurs quite
close to the black hole, at $\approx 2-3''$;\footnote{Asymptotically
the solutions in Figs. 1 and 2 approach that of Bondi accretion at
small radii (Bondi \& Hoyle 1944) and that of a thermally driven wind
at large radii (e.g., Parker 1960).} in 1D this separation occurs at a
stagnation point where $v = 0$.  The temperature near the stagnation
point, at $\sim 1-10''$, is set by the stellar wind velocity; for $v
\ll c_s$, the steady state solution to equation (3) is $c^2_s \approx
v_w^2/5$, which yields a characteristic temperature of $\approx 1$
keV, as is observed by \ch (Baganoff et al. 2003).

The density profile of the gas is shown in Fig. 2.  Also shown are two
\ch measurements (Baganoff et al. 2003); the first is an inferred
density of $\approx 27$ cm$^{-3}$ in the central 10'' from the diffuse
thermal emission in the Galactic Center.  The second is an inferred
density of $\approx 130$ cm$^{-3}$ in the central 1.5'' from the
extended X-ray source coincident with Sgr A* (discussed more below).
The agreement between the models and the \ch observations shows that,
although the spherically symmetric approximation has its limitations,
it captures the overall dynamics of hot gas in the Galactic Center
reasonably well.  It is worth stressing that in our model, the two \ch
observations probe gas with very different dynamics: the $\approx
10''$ observation probes the majority of the gas that is being driven
out of the central star cluster away from the black hole, while the
$\approx 1.5''$ observation probes the small fraction of the gas that
is gravitationally captured by the black hole.

Figure 3 shows the surface brightness profile within 3'' of Sgr A*
derived from the density and temperature profiles shown in Fig. 1-2.
Also shown is the surface brightness profile observed by \ch
coincident with Sgr A*, and that of a nearby point source (as an
indication of the \ch point spread function; see Baganoff et
al. 2003).  Aside from epochs when the source flares (Baganoff et
al. 2001; Porquet et al. 2003), the X-ray source coincident with Sgr
A* is significantly extended with respect to nearby point sources,
although there are uncertainties in the surface brightness profile of
Sgr A* because X-ray scattering by dust grains along the line of sight
can make the profile somewhat more extended (see Tan \& Draine 2003,
who estimate that dust scattering of an unresolved source can, at
most, account for $\sim 50 \%$ of the extended emission coincident
with Sgr A*).

The models presented here of thermal emission produced by gas
gravitationally captured by Sgr A* reproduce the extended X-ray source
reasonably well.  Naturally, models with steeper density profiles
(e.g., $\eta = 3$) produce steeper surface brightness profiles more in
accord with the \ch observations.  I have found it difficult to fully
reproduce the rapid drop in surface brightness that is observed.
Models with such a steep surface brightness profile either overpredict
the density by a factor of few or underpredict the temperature by a
comparable factor.  There are several possible explanations for this
discrepancy.  For example, it may be due to limitations of the
spherically symmetric dynamical model.  In addition, the timescale for
electrons and protons to come into thermal equilibrium with each other
is $\approx 200 \ n^{-1}_{100} \ T_1^{3/2}$ years, where $n_{100} =
n/100 \ {\rm cm^{-3}}$ and $T_1 = T_e/1 \ {\rm keV}$; this is longer
than the characteristic flow time at 1'', $R/v_w \approx 40$ years.
Thus it is possible that the electrons and protons do not have the
same temperature on the scales observed by \ch (if, e.g., they are
shock heated to different temperatures, as is typically observed in
supernova shocks).\footnote{The good agreement between the observed
electron temperature and the temperature expected from shocked stellar
winds suggests that this is not likely to be a very large effect; even
a factor of $\approx 1.5-2$ difference would, however, modify the
comparisons made here.} In this case the observed electron temperature
would not be a good proxy for the total pressure, as is assumed in the
models considered here.  In principle, this could be tested by
observing lines from ions such as iron, oxygen, or nitrogen, as is
done to probe the temperature structure of supernova remnants (e.g.,
Vink et al. 2003).  In practice this may not be possible because many
of the relevant lines are in the soft X-rays which are heavily
absorbed towards the Galactic Center, and because the lines will be
too narrow to be resolved by \ch (Astro-E2 may help with the latter
problem).


\section{Discussion}

The models presented here describe the dynamics of the hot gas
produced by shocked stellar winds in the Galactic Center, assuming for
simplicity spherical symmetry. They quantitatively account for the
observed level of diffuse X-ray emission in the central parsec,
predicting an electron density on $\approx 10''$ scales of $\approx
20-30$ cm$^{-3}$, in good agreement with \ch observations (Fig. 2).
This emission is produced by gas that is not bound to the black hole
and is being thermally driven out of the central star cluster in a
wind. This wind can have important dynamical effects on the
surrounding interstellar medium (e.g., Yusef-Zadeh \& Wardle 1993); it
may also be an important source of mass for the thermal X-ray emitting
'lobes' observed symmetrically around Sgr A* by \ch (Morris et
al. 2002).

In our models, a few percent of the mass supplied by stellar winds to
the central parsec is gravitationally captured by Sgr A*, implying an
accretion rate (at large radii) of $\approx 10^{-5} \mpy$.  As this
gas moves in towards the black hole it is compressed, resulting in an
increase in the gas density and X-ray surface brightness close to Sgr
A* that are in reasonable agreement with \ch observations (Fig. 2 \&
3).  I suggest that this agreement provides strong evidence that \ch
has directly observed gas being gravitationally captured by Sgr A*,
confirming one of the long-standing predictions of theoretical
accretion models (e.g., Melia 1992; Narayan et al. 1995).  There is
also evidence from the linear polarization of mm emission from Sgr A*
that the density close to the black hole is much less than a
straightforward extrapolation of the Bondi accretion rate to small
radii (e.g., Bower et al. 2003).  This is in accord with theoretical
predictions that very little of the gas captured at large radii
actually accretes onto the black hole (e.g., Blandford \& Begelman
1999).

\acknowledgements

I thank Fred Baganoff, Bruce Draine, and Jonathan Tan for useful
discussions; conversations with Anatoly Spitkovsky on numerical
methods for solving PDEs were particularly useful.  This work was
supported in part by NSF grant AST 0206006, NASA Grant NAG5-12043, and
an Alfred P. Sloan Fellowship.



\begin{figure}
\plotone{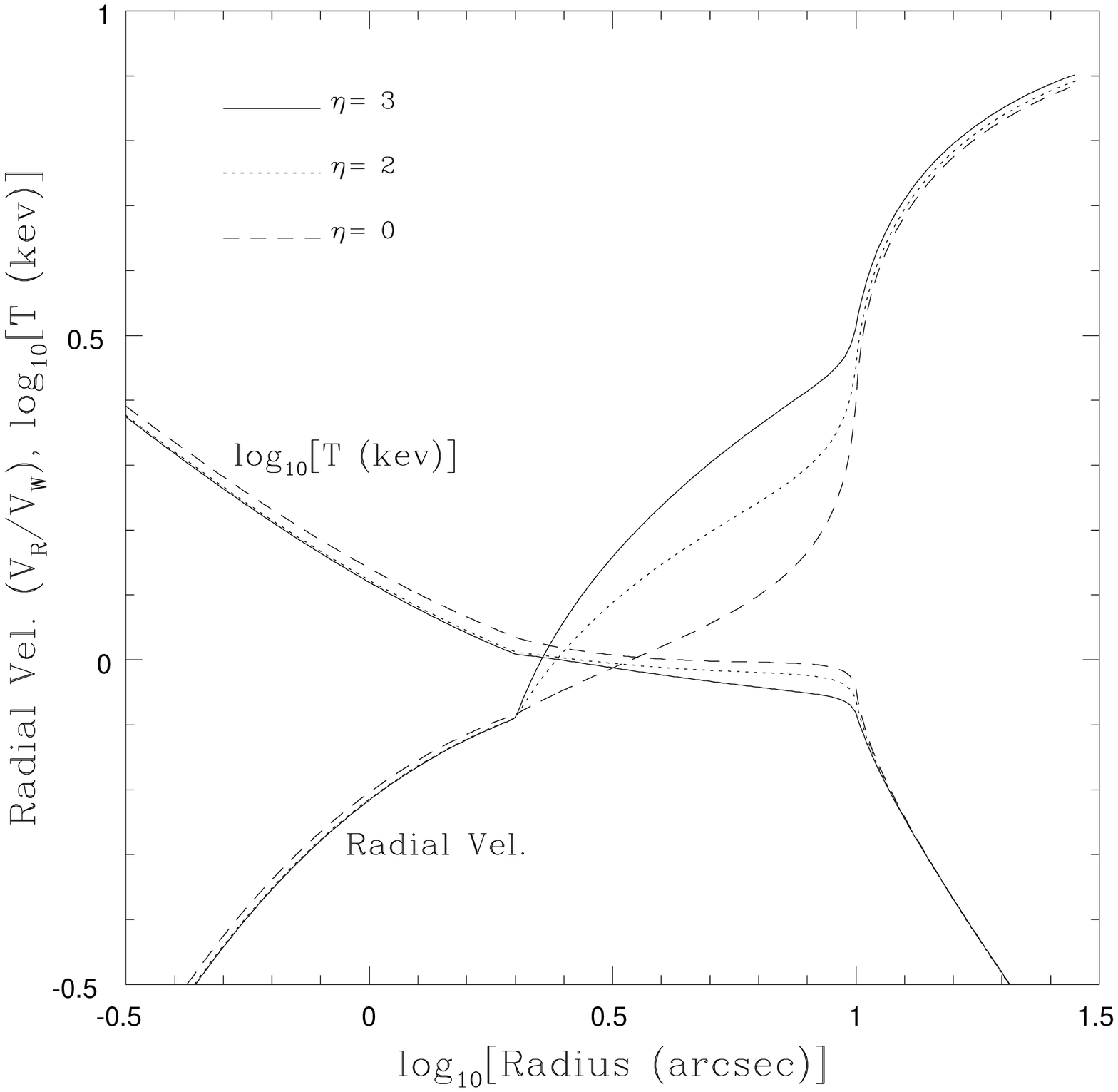}
\caption{Steady state radial velocity and gas temperature as a
function of distance from Sgr A*; the mass source term in the
continuity equation is $\propto r^{-\eta}$ for $r \ \epsilon \
[2'',10'']$.  Models for several values of $\eta$ are shown to
indicate the dependence of the results on the spatial distribution of
stellar mass loss (which is somewhat uncertain because of projection
effects and our spherically symmetric approximation).  The temperature
is the electron temperature calculated assuming $T_i = T_e$ and a mean
molecular weight of $\mu = 0.5$.}
\end{figure}

\begin{figure}
\plotone{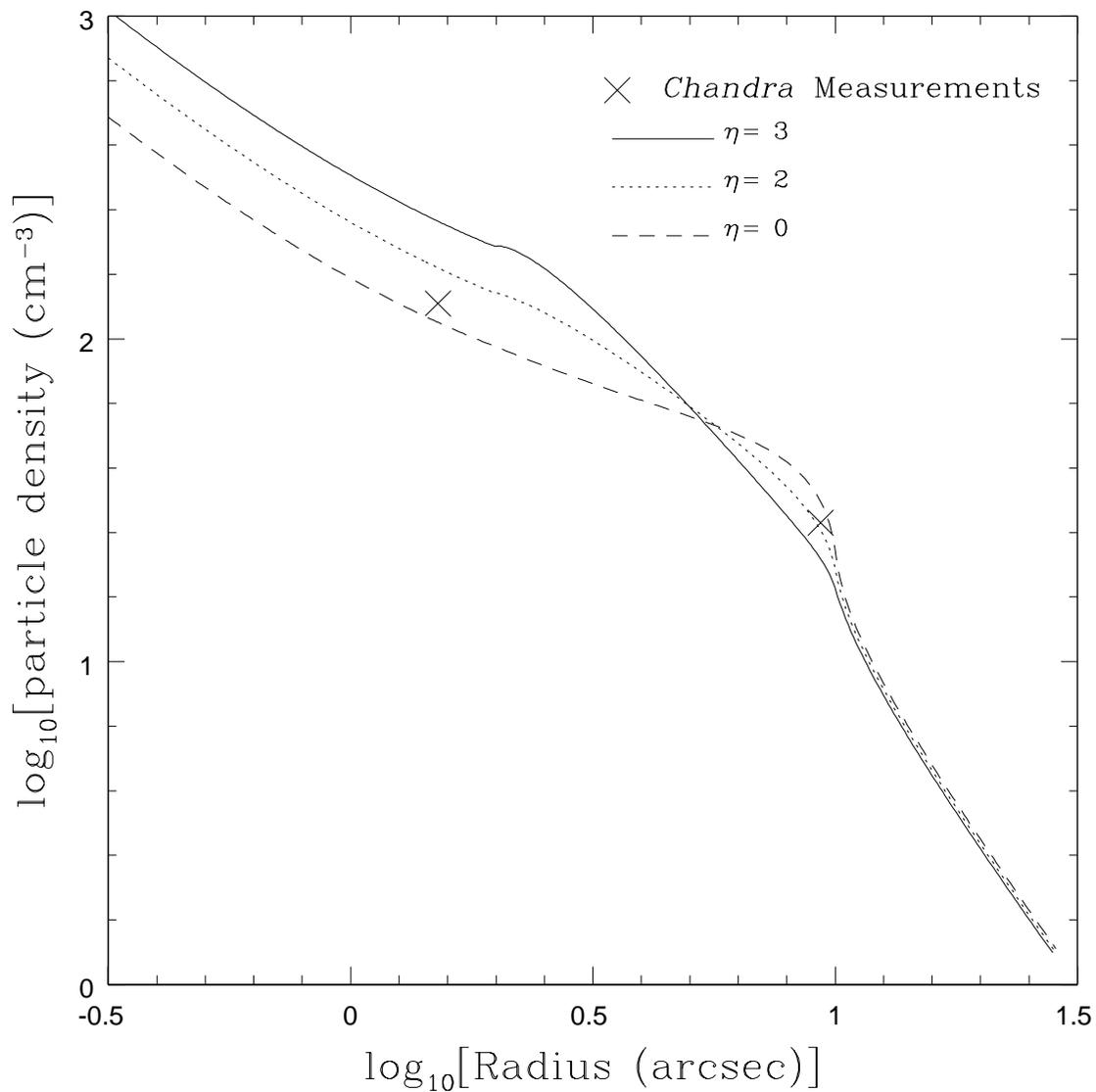}
\caption{Steady state gas density profiles for the same models as in
Fig. 1.  The data points are two \ch measurements of the electron
density from Baganoff et al. (2003).}
\end{figure}

\begin{figure}
\plotone{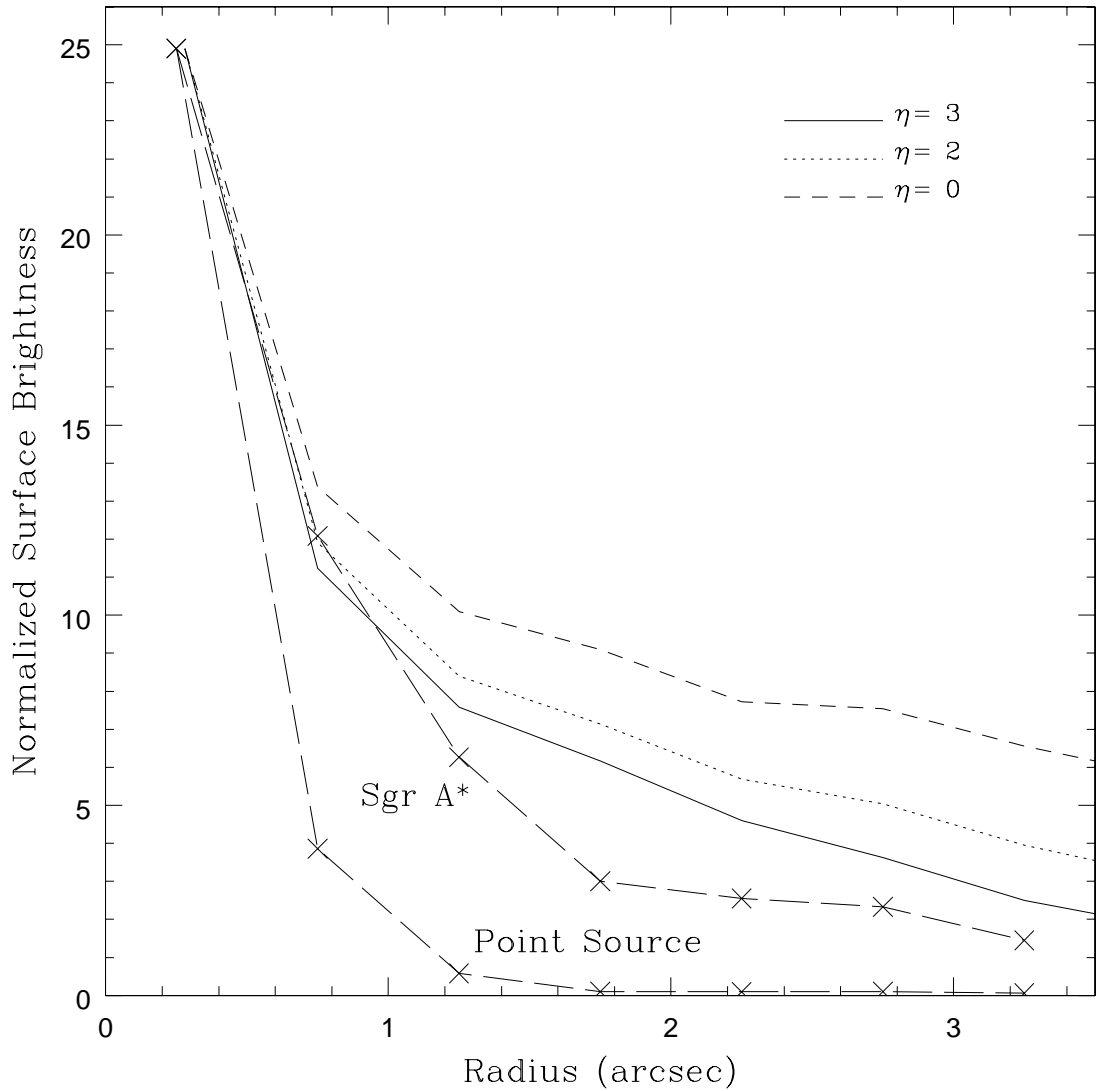}
\caption{X-ray surface brightness profiles (counts/pixel$^2$) in the
central 3'' around Sgr A*; the theoretical models use the density and
temperature profiles from Figs 1 \& 2 assuming thermal bremsstrahlung
to calculate the X-ray emission.  The profiles are normalized to have
the same value in the central pixel at 0.25''.  The data for Sgr A* in
non-flaring epochs, and the data for an X-ray point source, are taken
from Baganoff et al. (2003); no correction has been made for X-ray
scattering by dust grains (see Tan \& Draine 2003).  For Sgr A*, the
'background' subtracted off by Baganoff et al. (1.19 counts/pixel$^2$)
has been added back in because it represents diffuse emission modeled
here.}
\end{figure}

\end{document}